\documentstyle[eqsecnum,aps,prd,epsf]{revtex}

\def\msol{\ifmmode M_\odot\else$M_\odot$\fi}

\begin{document}



%

\bigskip
\bigskip
\bigskip

\begin{center}

{\Large \bf
Extreme Mass Ratio Binary: 
Radiation reaction and gravitational waveform
}

\bigskip

{\large Yasushi Mino
\footnote{Electronic address: mino@tapir.caltech.edu}}\\

\medskip

{\em mail code 130-33 Caltech Pasadena CA 91125 USA}\\

\medskip

\today

\end{center}

\bigskip

{\bf Abstract} \\

For a successful detection of gravitational waves by LISA,
it is essential to construct theoretical waveforms
in a reliable manner.
We discuss gravitational waves
from an extreme mass ratio binary system
which is expected to be a promising target of the LISA project.

The extreme mass ratio binary is a binary system of
a supermassive black hole and a stellar mass compact object.
As the supermassive black hole dominates 
the gravitational field of the system,
we suppose that the system might be
well approximated by a metric perturbation of a Kerr black hole.
We discuss a recent theoretical progress
in calculating the waveforms from such a system.

\section{Introduction} \label{sec:intro}

We consider a method to construct 
theoretical templates of gravitational waves LISA will detect. 
We are planning to operate LISA around a few years, 
and we may observe gravitational waves for this time scale. 
One of promising targets for LISA is 
a binary system of a supermassive black hole 
and an inspiralling compact object, 
and we discuss a method to calculate a gravitational waveform 
coming from such a system. 
Since the supermassive black hole dominates 
the gravitational field of the system, 
we expect that a linear metric perturbation of a Kerr black hole 
induced by a point particle 
might be a useful tool for this calculation. 

Because we already have a method
to calculate a metric perturbation of a given orbit of the particle,
it was widely believed that
the key issue here is to calculate a self-force, 
so that one can derive the orbit of the particle. 
We derived a formal expression of the self-force\cite{sf}
and we come to have a promising method to calculate
the self-force to the particle\cite{self}.
However, there was a concern on this approach
because of a gauge dependence of the self-force.

We discuss that the self-force is totally gauge dependent. 
By a special choice of the gauge condition, 
the self-force vanishes for the entire time interval 
where the linear metric perturbation is valid. 
(See Sec.\ref{sec:sf?}.) 
On the other hand, 
gravitational radiation reaction is observable
by asymptotic gravitational waves, 
and we expect that a physically reasonable self-force 
must describe this effect. 
Thus, the vanishing self-force suggests that
we need a substantial extension
of the theoretical basis in calculating the self-force, 
especially, of the gauge condition. 

We find a key idea to solve this problem is
an adiabatic approximation of the metric perturbation, 
which is different from a usual metric perturbation scheme. 
The idea of an adiabatic approximation of the metric 
changes our understanding of the orbital evolution
and the validity of the self-force was clarified for the first time.
We find that the class of gauge conditions
used in the radiation reaction formalism in Ref.\cite{RR}
is compatible with the adiabatic approximation
of the metric perturbation.
With a further choice of the gauge condition,
we find that the radiation reaction formula might describe
the orbit sufficiently long enough for the LISA project.

\section{Self-force? a myth?} \label{sec:sf?}

Gravitational radiation reaction is physically real
since we can define the gravitational flux
at the asymptotically flat region, 
and we expect to see this effect on the orbit
by a modulation of an observed gravitational waveform.
In order to predict this modulation,
we were motivated to calculate the self-force
for the belief that the self-force tells us
the gravitational radiation reaction effect on the orbit.
However, we find that the self-force could vanish 
consistently with a usual perturbation method 
irrespective to gravitational radiation reaction. 

For an approximation of an extreme mass ratio binary system,
we take the mass ratio of the binary $\mu$ as a small parameter
for the expansion.
In a usual perturbation scheme,
we expand the metric and the orbit 
from background quantities of the supermassive black hole as
\begin{eqnarray}
g_{\alpha\beta} &=& g^{kerr}_{\alpha\beta}+\mu h^{(1)}_{\alpha\beta}
+\mu^2 h^{(2)}_{\alpha\beta}+\cdots \,, \\
z^\alpha(\lambda) &=& z^{(kerr)\alpha}(\lambda)+\mu z^{(1)\alpha}(\lambda)
+\mu^2 z^{(2)\alpha}(\lambda)+\cdots \,,
\end{eqnarray}
where $g^{kerr}_{\alpha\beta}$ is a Kerr metric, 
and $z^{(kerr)\alpha}(\lambda)$ is a geodesic around it 
with $\lambda$ as an orbital parameter. 
For a valid perturbation, we assume
\begin{eqnarray}
&& O(1) \,\sim\, g^{bg}_{\alpha\beta}
\,>\, \mu h^{(1)}_{\alpha\beta}
\,>\, \mu^2 h^{(2)}_{\alpha\beta} \,>\, \cdots \,, \\
&& O(1) \,\sim\, z^{(bg)\alpha}(\lambda)
\,>\, \mu z^{(1)\alpha}(\lambda)
\,>\, \mu^2 z^{(2)\alpha}(\lambda) \,>\, \cdots \,,
\end{eqnarray}
and the perturbative equations are derived 
by expanding the Einstein equation with respect to $\mu$. 

One can see that this perturbation scheme
allows only a small deviation of the orbit 
from the background geodesic. 
Because the orbit will deviate 
from the background geodesic eventually 
by gravitational radiation reaction, 
this perturbation scheme is valid in a finite time interval. 

We consider a gauge transformation of the orbit 
in this finite time interval where the perturbation is valid. 
Because the deviation from a geodesic is $O(\mu)$,
one can always take a gauge transformation
to eliminate the deviation during this whole time interval, 
i.e. $\mu z^{(1)\alpha}=\mu^2 z^{(2)\alpha}=0$, 
thus, the self-force totally vanishes 
consistently with the usual perturbation scheme. 

\section{Adiabatic Metric Perturbation} \label{sec:ad}

The reason for this gauge problem of the self-force is 
that the perturbation is valid only in a short time scale, 
so that we can only describe a small deviation of the orbit 
from the background geodesic. 
In order to solve this problem, we propose an extension 
of the perturbation scheme. 

Bound geodesics around a Kerr black hole 
with a convenient orbital parameter $\lambda$ 
are characterized 
by three constants of motion ${\cal E}^a,\,a=E,L,C$,
and four phase constants $\lambda^b,\,b=r,\theta,\,C^c,\,c=t,\phi$ 
as in Ref.\cite{RR}, 
and we denote them collectively
by $\gamma = \{{\cal E}^a,\lambda^b,C^c\}$.
From the result of Ref.\cite{RR},
the orbital deviation becomes $O(1)$ when $\lambda \sim O(\mu^{-1/2})$.
This time scale is called the dephasing time.
From the discussion in Sec.\ref{sec:sf?},
the usual perturbation scheme is valid only in this time scale.
Our purpose here is to extend this time scale of validity
by modifying the metric perturbation scheme.

In the usual perturbation scheme, 
the linear metric perturbation is induced by a geodesic, 
and we can write it as a function of $\gamma$, 
$h_{\alpha\beta}(x)=h_{\alpha\beta}(\gamma;x)$.
Here we use a class of gauge conditions
where the metric perturbation is derived by the Green function 
in the Boyer-Lindquist coordinates as 
\begin{eqnarray}
h_{\alpha\beta}(x) &=& \int dx' 
G_{\alpha\beta\,\alpha'\beta}(x,x')T^{\alpha'\beta'}(x') \,, \\ 
G_{\alpha\beta\,\alpha'\beta'}(x,x') &=& 
G_{\alpha\beta\,\alpha'\beta'}(t-t',\phi-\phi',r,r',\theta,\theta')
\label{eq:gau} \,.
\end{eqnarray}
We assume an adiabatic approximation to the orbit as in Ref.\cite{RR}, 
and consider the evolution of "geodesic constants" 
by the effect of gravitational radiation reaction. 
We write the "geodesic constants"
as functions of the orbital parameter, $\gamma(\lambda)$.
We consider to extend this idea to the metric perturbation.
We foliate the spacetime by spacelike hypersurfaces. 
Using the fact that the foliation surfaces intersect with the orbit,
we define the foliation function $f(x)$ by the orbital parameter
at the intersection of the orbit and the surfaces as
$f(z(\lambda))=\lambda$.
We define 
the adiabatic metric perturbation on the foliation surface $f(x)$ 
by the linear metric perturbation induced by the geodesic $\gamma(f)$ 
as 
\begin{eqnarray}
h^{ad}_{\alpha\beta}(x) &=& h_{\alpha\beta}(\gamma(f);x)
\label{eq:adm} \,.
\end{eqnarray}

In order to see the validity of the adiabatic metric perturbation,
we operate the linearized Einstein operator and we have
\begin{eqnarray}
G^{(1)}_{\alpha\beta}[h^{ad}] &=& T_{\alpha\beta}[\gamma(f)]
+\Lambda^{(1)}_{\alpha\beta}[h^{ad}] \,, \label{eq:adme}
\end{eqnarray}
where an extra term $\Lambda^{(1)}_{\alpha\beta}$ appears
because the adiabatic metric perturbation is not induced by a geodesic.
It is notable that $T_{\alpha\beta}[\gamma(f)]$
is the stress-energy tensor of a point particle
moving along the orbit 
with the effect of gravitational radiation reaction.
The adiabatic metric perturbation solves
the Einstein equation to the accuracy of $O(\mu)$
as long as $O(\mu)>\Lambda^{(1)}_{\alpha\beta}$ holds.
Since $\Lambda^{(1)}_{\alpha\beta}$ depends
on the $\lambda$ derivatives of the orbital "constants"
$\gamma(\lambda)$,
the validity of the adiabatic metric perturbation
depends on how the orbit evolves.
We find that, under the gauge condition (\ref{eq:gau}),
the adiabatic metric perturbation is valid
in the radiation reaction time scale $O(\mu^{-1})>\lambda$.

Because the time scale of validity is much longer
than that of the linear perturbation,
the orbital deviation could be $O(1)$ 
in this metric perturbation scheme 
and one cannot eliminate the orbital deviation from the geodesic 
by a gauge transformation,
thus, the self-force cannot vanish by a gauge transformation 
and is physically meaningful 
to predict the orbit in this class of gauge conditions.

\section{Adiabatic Evolution of the Orbit} \label{sec:rr}

Ref.\cite{RR} finds that
the infinite time averaged self-force
acting on the "constants of motion", $<(d/d\lambda){\cal E}^a>$,
can be derived by the radiative metric perturbation
in a gauge invariant way
by a simple symmetry property of a Kerr spacetime.
It also find that $<(d/d\lambda){\cal E}^a>$ makes
a dominant contribution to the orbital evolution
by a perturbative analysis of the orbit.

The adiabatic metric perturbation is valid
in the radiation reaction time scale
and, in this time scale,
the orbital evolution is non perturbative.
As for the "constants of motion",
one can still deal with the evolution by a perturbation,
and we find that $<(d/d\lambda){\cal E}^a>$ dominantly determines
the evolution of them.
The subdominant part of the "constants of motion" is $O(\mu)$.
The evolution of the "phase constants" becomes
non perturbative beyond the dephasing time,
however, we find the perturbative results in Ref.\cite{RR}
are qualitatively correct.
The dominant part of their evolution is described
only by $<(d/d\lambda){\cal E}^a>$,
and the subdominant part behaves as $O(\mu \lambda)$.

Because the subdominant part of the evolution of "constants of motion"
is smaller than $O(1)$ in the radiation reaction time scale,
one can find a gauge condition which can totally eliminate this part.
We call this gauge condition by the radiation reaction gauge,
and, in this gauge, the self-force can be written only 
by the infinite time averaged self-force as
\begin{eqnarray}
(d/d\lambda){\cal E}^a \,=\, <(d/d\lambda){\cal E}^a> \,.
\end{eqnarray}

The self-force is expected to have the conservative part
as well as the radiative part,
however, in this gauge condition,
the conservative part is integrated out
and appears to be the small shift of initial values. 
We also note that the radiation reaction gauge condition
applies to an orbit of a spinning particle.
It is known that the orbit of the spinning test particle
slightly deviates from a geodesic\cite{papa}
because of the coupling of its spin and the background curvature.
This effect of the deviation can also be absorbed 
by a small change of the initial values
by the radiation reaction gauge condition.

\section{Self-Force Calculation vs Radiation Reaction Formula}

Here we try to compare these two approaches to calculate the orbit;
the regularization calculation of the self-force\cite{self}
and the radiation reaction formula discussed in Sec.\ref{sec:rr}.
There could be two issues for comparison;\\
{\bf 1)}
{\it Theoretical issue; whether the resulting orbit
is physically acceptable or not} \\
{\bf 2)}
{\it Practical issue; whether the approach is practically available
for generating templates for the LISA project} \\

Gravitational radiation reaction is a physically real effect.
Because the momentum flux carried by gravitational waves 
is well defined,
it is reasonable to expect that 
the orbital evolution must be consistent to this effect.
There is a wide-spread belief that
the self-force includes the effect of gravitational radiation reaction,
however, this is not trivial in the usual perturbation scheme 
because of the gauge freedom as we argue in Sec.\ref{sec:sf?}.
We consider that the regularization calculation of the self-force\cite{self} 
is not guaranteed to include this radiation reaction effect in general.
On the other hand, the radiation reaction formula 
uses a certain class of gauge conditions 
as we argue in Sec.\ref{sec:ad}, 
and the orbital evolution is described consistently 
with asymptotic gravitational radiation. 

For a practical issue,
an idea is proposed to calculate the metric perturbation 
in a Kerr black hole 
available for the self-force calculation\cite{amos},
however, it is not coded yet, 
and a more crucial problem is 
whether the resulting self-force is consistent 
with gravitational radiation reaction.
Even if it is possible, the numerical convergence 
of the self-force is not optimal 
since it is a point splitting regularization. 
On the other hand,
the radiation reaction formula is, in the end,
just an extension of the balance formula,
and various numerical codes of the balance formula
were already made successfully\cite{pp}. 
There is also known a semi-analytic technique
which may substantially increase the efficiency of the code\cite{mano}.
In this approach, we have the most optimal convergence
because the regularization is done in a quite natural manner,
and we consider it helpful to have an efficient calculation.

\section*{Acknowledgement}

YM thanks to Prof. Kip Thorne and Prof. Sterl Phinney
for encouragement.
YM was supported by NASA grant NAG5-12834
and NASA-ATP grant NNG04GK98G at CalTech.




\end{document}